\documentclass[aps,10pt,prb,twocolumn,groupedaddress,showpacs]{revtex4}

\usepackage{graphicx}
\usepackage[caption=false]{subfig}
\usepackage{amsmath}
\usepackage{bbold}
\usepackage{latexsym}

\begin{document}


\title{Exact zero modes and decoherence in systems of interacting Majorana fermions}


\author{Guang Yang and D. E. Feldman}
\affiliation{Department of Physics, Brown University, Providence, Rhode Island 02912, USA}


\date{\today}

\begin{abstract}
Majorana fermions often coexist with other low-energy fermionic degrees of freedom. In such situation, topological quantum computation requires the use of fermionic zero modes of a many-body system. We classify all such modes for interacting fermions and show how to select the mode that maximizes the decoherence time. We find that in a typical interacting system the maximal decoherence time is within one order of magnitude from the decoherence time of a qbit based on the local part of the fermion parity operator.
\end{abstract}



\maketitle



\section{Introduction \label{sec:intro}}

The beautiful idea of topological quantum computation \cite{kitaev,nayak} offers a conceptually simple and straightforward approach to quantum information processing: Logical operations can be performed by braiding topological excitations and the memory remains protected from errors as long as the ground state manifold is separated from the excitations by a sufficient gap. Several systems are expected to host non-Abelian particles which can be used to implement topological quantum computing. In particular, much attention has focused on  fractional quantum Hall states in the second Landau level\cite{nayak}, $p$-wave superconductors\cite{read,ivanov},  and heterostructures of superconductors and topological insulators\cite{fu}. In a majority of those systems, topological excitations are bound states of Majorana fermions. Majorana fermions are insufficient for the universal quantum computation 
\cite{nayak,freedman2002a,freedman2002b} but they do provide a route to topologically protected memory.

A Majorana fermion can be thought of as a half of a complex fermion. Thus, a system of two distant Majorana fermions $\gamma_1$ and $\gamma_2$ possesses two degenerate quantum states which differ by their fermion parity. 
Those two states can be used to form a qbit (strictly speaking, one needs four Majorana fermions but this will not be important below). No local operators that affect parity can be constructed from $\gamma_1$ and $\gamma_2$ and hence the qbit enjoys topological protection.
The situation changes, if other low-energy fermionic excitations are present. For example, a Majorana fermion in the core of a superconducting vortex is separated from other excitations in the vortex by a tiny minigap
\cite{caroli,kopnin}. Within the mean-field approximation those excitations do not interact with Majorana fermions but corrections to the mean-field theory are always present and can be significant and comparable to the minigap \cite{footnote1}. As a result, the parity of the Majorana qbit no longer conserves.

An elegant way around this problem was proposed by Akhmerov in Ref. \onlinecite{akhmerov}. The topological charge of a closed subsystem that contains a Majorana fermion must always conserve. As a consequence, the local part $\Gamma$ of the fermion parity operator can be used in place of the original Majorana fermion to obtain a protected qbit. The form of the local parity operator $\Gamma$ does not depend on any details of the closed system except the number  of its degrees of freedom. Thus, the same solution works for any interaction, weak or strong, and even time-dependent Hamiltonians.
The only weakness of the proposal is related to the fact that a Majorana qbit is never an ideal closed system \cite{GS2011,budich,scheurer,hassler}. For example, it can exchange fermions with metallic gates used to control the system
\cite{schmidt}. Another problem comes from quasiparticle poisoning \cite{p1,p2,p3,p4,p5,p6,p7,p8,p9,p10}. 
In the ideal equilibrium limit, the number of bulk excitations scales as $\exp(-\Delta/T)$, where $\Delta$ is the energy gap, and hence is vanishingly small at low temperatures. However, in real low-temperature superconductors a nonequilibrium quasiparticle population is present and may considerably limit the qbit lifetime. The local parity $\Gamma$ is much more vulnerable to these and other decoherence mechanisms than an individual Majorana fermion.

Ref. \onlinecite{GS2012} has argued that other fermionic zero modes can be used to build a qbit  with a longer lifetime than in Akhmerov's proposal. An appropriate zero mode has been identified for a quadratic Hamiltonian. Its use  significantly increases the decoherence time indeed. That observation could be anticipated from the fact that quadratic Hamiltonians correspond to the mean-field approximation. One can expect similar behavior in the limit of weak interactions. What happens beyond that limit remains an open question. We answer that question below. We find simple analytic expressions for all zero modes of a general fermionic Hamiltonian, use those expressions to estimate the maximum decoherence time, and give an algorithm for designing a qbit with the longest lifetime. At weak interaction, in a system of $2N+2$ Majorana fermions, the lifetime $\tau$ can be increased to about $2N\tau_\Gamma$, where $\tau_\Gamma$ is the decoherence time of the qbit, based on the local parity operator $\Gamma$. At strong interaction the gain in the lifetime is less spectacular: $\tau \lesssim 10\tau_\Gamma$.

This paper is organized as follows. As a warming-up exercise, in Section II, we consider the simplest case of one Majorana and one complex fermion. In Section III we classify all zero modes of a general interacting fermionic Hamiltonian and describe those modes which can be used in a qbit. A general expression for their decoherence time is obtained in Section IV. We estimate the maximal decoherence time for strongly interacting systems in Section V.
Sections II-V focus on systems made of fermions only. This is the main question addressed in this article. What happens in the presence of additional bosonic modes, such as phonons, is briefly discussed  in Section VI.
We summarize our results in Section VII. Four Appendices contain technical details.

\section{One real and one complex fermion}

This case is easy and always reduces to the mean-field limit considered in Ref. \onlinecite{GS2012}. We will generalize for an arbitrary number of degrees of freedom in subsequent sections.

One complex fermion is equivalent to two real fermions: $c^\dagger=(\gamma_1+i\gamma_2)/2$. Thus, it is sufficient to study the problem with three Majorana fermions $\gamma_0,\gamma_1,\gamma_2$ with the anticommutation relations
$\{\gamma_i,\gamma_j\}=2\delta_{ij}$.

Let $f$ be another Majorana operator, localized far away from the subsystem, where $\gamma_i$ live.
Consider a Majorana operator $F$, constructed from $\gamma_i$. Since $F$ obeys the Fermi statistics and anticommutes with $f$, it is a polynomial of an odd degree as a function of $\gamma_i$, $i=0,1,2$. $F$ must also be Hermitian and satisfy the Majorana condition $F^2=1$.
 Then the operator $iFf$ has two parity eigenvalues $\pm 1$ which can be used to store quantum information. The information is preserved as long as the parity does not change. In a closed system, the parity conserves indefinitely as long as  $F$ commutes with the Hamiltonian, i.e., is a zero mode. 
Thus, as the first step, we classify fermionic zero modes. 

The Hamiltonian that controls the $\gamma_i$ degrees of freedom is a Bose-operator and reduces to a sum of products of even numbers of Majorana fermions. Since $\gamma_i^2=1$, the Hamiltonian is quadratic:

\begin{equation}
\label{1}
H=2i(a_0\gamma_1\gamma_2+a_1\gamma_2\gamma_0+a_2\gamma_0\gamma_1)=i\sum_{ij}A_{ij}\gamma_i\gamma_j,
\end{equation}
where $A_{ij}$ is a skew-symmetric matrix. Any three-dimensional skew symmetric matrix can be reduced to a block diagonal form

\begin{equation}
\label{2}
\tilde A=\left( \begin{array}{ccc}
0 & 0 & 0 \\
0 & 0 & a \\
0 & -a & 0 \end{array} \right),
\end{equation}
by an orthogonal transformation $\tilde\gamma_i=\sum_j O_{ij}\gamma_j$, where $\tilde\gamma_i$ is a new set of Majorana operators. Hence, the Hamiltonian can be rewritten as $H=2ia\tilde\gamma_1\tilde\gamma_2$. In order to identify fermionic zero modes $F$ we need to compute the commutator $[H,F]=0$. We find that all (Hermitian) zero modes $F$ reduce to linear combinations  $F=\alpha\tilde\gamma_0+\beta\Gamma$ with real coefficients $\alpha$ and $\beta$, where $\Gamma=i\tilde\gamma_0\tilde\gamma_1\tilde\gamma_2$ is the local parity operator. Since $F$ is a Majorana fermion, $F^2=1$. Therefore there are only two possibilities for $F$ (up to an overall sign): $F=\pm\tilde\gamma_0$ or $F=\pm\Gamma$. The first choice corresponds to Ref. \onlinecite{GS2012} and the second choice coincides with Akhmerov's proposal \cite{akhmerov}.

Both choices would work equally well in a closed system. We now wish to include decoherence effects due to external degrees of freedom. As argued in Ref. \onlinecite{GS2012}, the decoherence time $\tau_\Gamma$ for $F=\Gamma$ is shorter than the decoherence time $\tau_0$ for $F=\tilde\gamma_0$. 
We review the estimate of the lifetimes below. We generalize it for larger systems in subsequent sections.

The lifetime depends on the details of the interaction with the bath. Thus, for the most general case, only a crude estimate can be obtained for the ratio of $\tau_0$ and $\tau_\Gamma$. Our estimate is based on a simple model of a bath as a system of noninteracting electrons $c_{k,i}$ with the Hamiltonian $H_b=\sum_{k,i}\epsilon_{k,i}c^\dagger_{k,i}c_{k,i}$. The index $k$ is continuous and $i$ labels various discrete degrees of freedom. The calculations simplify slightly, if we assume that each operator $\tilde\gamma_i$ couples only with the bath fermions labeled by the same index $i$. Thus, the interaction with the bath is described by the Hamiltonian $H_I=\sum_{i,k} M_{k,i}(t)
\tilde\gamma_i c_{k,i}+ H.c.$  

We will estimate the lifetime $\tau$ of a qbit by computing how long it takes until the bath flips the sign of the parity eigenvalue $iFf=\pm 1$. A similar estimate can be extracted from the time dependence of the parity correlation function $C(s)=\langle iF(t=0)f iF(t=s) f\rangle$ and the condition $C(\tau)\sim C(0)/e=e^{-1}$. Both approaches also lead to similar results in a system with an arbitrary number of Majorana fermions.

Let the system be in an initial state $|0\rangle$ with a definite parity of the qbit $q(t=0)=\langle 0|iFf|0\rangle$. $q(0)$ can be set equal to one without loss of generality. In order to compute the qbit lifetime we need to estimate the time-dependence of $q(t)=\langle\psi(t)|iFf|\psi(t)\rangle$, where $\psi(t)$ is the time-dependent wave function with the initial condition $|\psi(0)\rangle=|0\rangle$.
Let us find the probability of the transition  to the state
$|\psi_q\rangle |\psi_b\rangle$ during the time interval $t$, where $|\psi_q\rangle$ is a wave function of the qbit and $|\psi_b\rangle$ a state of the bath. The probability is given by Fermi's golden rule, $p\sim t |\langle \psi_q|\langle \psi_b| H_{I,\omega}
|0\rangle|^2$, where $H_{I,\omega}$ is the Fourier harmonic of the interaction at the frequency $\omega$, determined by the energy conservation: $\hbar\omega=E_{\rm final}-E_{\rm initial}$. 
The matrix element in the above formula is a linear combination of the products of the matrix elements $\langle \psi_q|\tilde\gamma_i|0\rangle$ and matrix elements of the fermion operators in the bath.
The relevant matrix elements and densities of states 
depend on the details of the physical realization.
 For the sake of a general estimate, we use the simplest assumption that the probability to find the qbit in a common energy and total parity eigenstate $|\psi_q\rangle$ is $P_q=A t \sum_i |\langle \psi_q|\tilde\gamma_i|0\rangle|^2$, where $A$ does not depend on $|\psi_q\rangle$. It is easy to compute $q(t)$ now:

\begin{align}
\label{3}
q(t)=1-At\sum_{i,\psi_q}|\langle \psi_q|\tilde\gamma_i|0\rangle|^2+ 
At\sum_{i,\psi_q}\langle\psi_q|iFf|\psi_q\rangle | \langle \psi_q|\tilde\gamma_i|0\rangle|^2 \nonumber\\
=1-At\sum_i[\langle 0|\tilde\gamma_i^2|0\rangle- 
\langle 0|\tilde\gamma_i iFf \tilde\gamma_i|0\rangle],
\end{align}
where it is legitimate to include $|\psi_q\rangle=|0\rangle$ in the sums in the first line.
By setting $F=\tilde\gamma_0$ we get $q(t)=q_0(t)=1-2At$ at $t\ll 1/A$. By setting $F=\Gamma$ we find $q(t)=q_\Gamma(t)=1-6At$. This shows that $\tau_0=3\tau_\Gamma\sim 1/[2A]$.

The above results crucially depend on the existence of a linear zero mode in any system of three Majorana particles. Such modes do not exist \cite{wilczek} in a generic system with more than three Majorana fermions. We address  the structure of the zero modes for an arbitrary number of the degrees of freedom  in the next section.

\section{Zero modes}

We consider a system of an arbitrary odd number $2N+1$ of  interacting Majorana fermions $\gamma_0,\dots,\gamma_{2N}$. This is equivalent to a system of one Majorana and $N$ complex fermions.
We first neglect the interaction with the bath. Then the Hamiltonian $H$ is a linear combination of various products of even numbers of Majorana operators.

As the discussion in the previous section shows, in order to construct a qbit, we need to identify a fermionic zero mode operator $F$, constructed from $\gamma_i$, such that $F=F^\dagger$, $[H,F]=0$ and $F^2=1$.
Then quantum information can be encoded in the total parity operator $iFf$, where $f$  is  a Majorana fermion far away from $\gamma_i$.
We start with finding all fermionic zero modes $O$, commuting with $H$, and impose the Majorana condition $O^2=1$ later.  

Obviously, $\Gamma=i^{N}\Pi_{i=0}^{2N}\gamma_i$ is Hermitian and commutes with any Hamiltonian $H$. The operator $H$ also commutes with itself. Thus, any operator of the form 

\begin{equation}
\label{add1}
O=P(H)\Gamma,
\end{equation}
where $P(x)$ is a polynomial,
is a fermionic zero mode. We show in Appendix B that there are no other fermionic zero modes for a generic Hamiltonian $H$.
We also find that for a generic Hamiltonian, there are exactly $2^N$ linear independent integrals of motion of the above form. Note that the number $2^N$ of fermionic zero modes has been identified in Ref. \onlinecite{GS2012} for non-interacting Hamiltonians and in the case of infinitesimal interactions. On the other hand, we establish a general result.

The number of the linear independent zero modes is easy to understand. We first note that any fermionic zero mode can be obtained by multiplying a bosonic zero mode by $\Gamma$. This establishes a one-to-one correspondence between bosonic and fermionic zero modes. Thus, it is sufficient to establish that there are exactly $2^N$ linear independent bosonic zero modes. For this end, we notice that the Hamiltonian acts in the Hilbert space of dimension $2^{N+1}$, defined by the $(2N+2)$ Majorana operators $\gamma_k$ and $f$. There are $2^N$ states of even parity $i\Gamma f$ and $2^N$ states of odd parity $i\Gamma f$ in the Hilbert space. The two subspaces are connected by the operator $\Gamma$. The Hamiltonian and all bosonic zero modes commute with $\Gamma$ as well as with the parity operator $i\Gamma f$. Hence, they can be represented in the form of block operators with two identical blocks in the even and odd subspaces. This means, in turn, that it is sufficient to classify bosonic zero modes of the restriction of $H$ to the even subspace. After the diagonalization of the Hamiltonian in that subspace, we obtain a diagonal matrix of size $2^N$.
Clearly, such matrix commutes with at least $2^N$ linear independent Hermitian operators that preserve the parity $i\Gamma f$ and commute with $\Gamma$, i.e., bosonic zero modes. Moreover, there are exactly $2^N$ linear independent modes in the generic case of the Hamiltonian without degenerate eigenvalues.

Appendix B contains a different proof, based on an explicit construction of the zero modes for particular Hamiltonians.

It is also easy to see that for a generic Hamiltonian,  $2^N$ linear independent fermionic zero modes can be selected in the form $O_k=H^k\Gamma$, $k=0,\dots,2^N-1$. Indeed, if those modes were linear dependent then one could find a polynomial $P(x)$ of a degree $n<2^N$ such that $P(H)=0$. Thus, $P(E_k)=0$ for all eigenvalues $E_k$ of the Hamiltonian. However, in a generic situation there are $2^N$ different energy levels $E_k$ in contradiction with the fundamental theorem of algebra. Hence, the modes $O_k$ are linear independent. Any other fermionic zero mode is a linear combination of the modes $O_k$, i.e., satisfies Eq. (\ref{add1}), where the degree of the polynomial $P(H)$ is less than $2^N$. 

Not all zero modes $O=P(H)\Gamma$, ${\rm deg~}[P(x)]<2^N$, are suitable to build a qbit. We need to impose the condition 

\begin{equation}
\label{normalization}
O^2=1.
\end{equation}
 This is equivalent to $P^2(H)=1$. In turn, the former condition simplifies to $P(E_k)=\pm 1$ where $E_k$, $k=1,\dots,N_E=2^N$, are the eigenenergies of $H$. All allowed polynomials $P$ can be written 
in terms of $E_k$ with the use of standard interpolation formulas, e.g.,

\begin{equation}
\label{interpolation}
P(x)=\sum_k P(E_k)\Pi_{{n=1}}^{k-1}\frac{x-E_n}{E_k-E_n}\Pi_{{n=k+1}}^{N_E}\frac{x-E_n}{E_k-E_n}.
\end{equation}
The number of the possible choices of $P$ depends on the number $N_E=2^N$ of the energy levels of $H$ and equals $2^{N_E}$. This set is made of $2^{N_E-1}$ pairs of opposite polynomials $P$ and $-P$. Thus, there are $2^{N_E-1}$ ways to build a topological qbit.

The goal of the next two sections is to estimate the maximal lifetime for such qbits in the presence of a bath.

\section{Dephasing}

In this section we derive a general formula for the lifetime of a qbit. We will use it to estimate the maximal possible lifetime in  a system of $2N+1$ interacting Majorana modes in the next section. The exact value of the lifetime depends on many details of the Hamiltonian and cannot be computed in a general situation. Thus, we limit ourself to an estimate based on a simple model in the spirit of Section II. 

Our model Hamiltonian is the sum of three pieces, $H+H_b+H_I$, where $H$ describes the Majorana degrees of freedom, $H_b=\sum_{k,i}\epsilon_{k,i}c^\dagger_{k,i}c_{k,i}$ is the bath Hamiltonian, and $H_I=\sum_{i,k} M_{k,i}(t)
\gamma_i c_{k,i}+ H.c.$  describes the interaction of the qbit with the bath. Quantum information is encoded in the total parity operator $P=iFf$, where $F$ is a fermionic zero mode (Section III) and $f$ is a Majorana fermion, located far away from the Majorana fermions $\gamma_0,\dots,\gamma_{2N}$. As discussed above, $F$ is a polynomial of $\gamma$'s: 

\begin{equation}
\label{add2}
F=\sum_{n=0}^{N} i^n \sum_{\{k_l\}} a_{k_1,\dots,k_{2n+1}}\Pi_{s=1}^{2n+1}\gamma_{k_s},
\end{equation}
where $k_1<k_2<\dots<k_{2n+1}$ and $\{k_l\}$ is the shorthand for the set of the indices $k_1,\dots,k_{2n+1}$.
The fact that $F$ is Hermitian implies that the coefficients $a_{k_1,\dots,k_{2n+1}}$ are real. The normalization condition (\ref{normalization}) means that

\begin{equation}
\label{add3} 
\frac{{\rm Tr} F^2}{2^{N+1}}=\sum a^2_{\{k_l\}}=1.
\end{equation}

In what follows we ignore the interaction between $f$ and the bath and only compute the lifetime due to the interaction of the bath with the Majorana modes $\gamma_k$. Assuming that the physics is similar in the  regions, where $F$ and $f$ are localized, one can expect that the interaction of $f$ with the bath cuts the lifetime in half. 

We assume that initially the qbit is in a common eigenstate $|q_k\rangle$ of the total parity operator $P$ and the Hamiltonian $H$, $P|q_k\rangle=\pm |q_k\rangle$. 
The $2^{N+1}$ states $|q_k\rangle$ form an orthonormal basis in the Hilbert space on which $P$ and $F$ act.
Due to the interaction with the bath, the average $q(t)=\langle P(t)\rangle$ depends on time and eventually approaches 0. Since $P^2=1$ for any choice of $|q_k\rangle$, a convenient definition of the relaxation time is

\begin{equation}
\label{4}
\tau=\frac{2P^2(0)}{-\frac{d}{dt}\overline{\langle P(t)\rangle^2}\bigg\rvert_{t=0}}=-\frac{2}{\frac{d}{dt}\overline{\langle P(t)\rangle^2}\bigg\rvert_{t=0}},
\end{equation}
where the bar denotes the average with respect to all possible choices of $|q_k\rangle$ and the angular brackets denote the average with respect to the time-dependent wave function.

In order to compute  $q(t)=\langle P(t)\rangle$ we need to know the density matrix of the qbit at the time $t$. 
This reduces to the calculation of the transition probabilities from the initial state to all other states $|q_i\rangle$.
The probabilities can be computed with Fermi's golden rule and depend on many features of the system. Following Section II, we make the simplest assumption that the transition probability between the states $|q_i\rangle$ and $|q_j\rangle$, $i\ne j$, has the form
$P_{i\rightarrow j}=At\sum_k|\langle q_i |\gamma_k| q_j\rangle|^2$, where $A$ does not depend on $i$ and $j$. Then

\begin{align}
\label{5}
q(t)=q(0)[1-\sum_{i,j}At|\langle q_i |\gamma_j| q_k\rangle|^2]\nonumber\\
+\sum_{i,j}At\langle q_k |\gamma_j| q_i\rangle\langle q_i|P|q_i\rangle\langle q_i |\gamma_j| q_k\rangle\nonumber\\=
q(0)[1-(2N+1)At]+At\sum_j\langle q_k|\gamma_j P \gamma_j|q_k\rangle=\langle q_k| P_t | q_k\rangle,
\end{align}
where we include $i=k$ in the sums, use the fact that $P$ is diagonal in the $|q_i\rangle$ basis, and set  $P_t=iQ_t f$ with

\begin{equation}
\label{6}
Q_t=\sum_{n=0}^{N} [1-2(2n+1)At] i^n \sum_{\{k_l\}} a_{k_1,\dots,k_{2n+1}}\Pi_{s=1}^{2n+1}\gamma_{k_s}.
\end{equation}
Note that

\begin{align}
\label{7}
\frac{d}{dt}\overline{\langle P(t)\rangle^2}\bigg\rvert_{t=0}=2\overline{\langle P(t)\rangle \frac{d\langle P(t)\rangle}{dt}}\bigg\rvert_{t=0}\nonumber\\
=2\frac{{\rm Tr}P_{t}\frac{dP_{t}}{dt}}{2^{N+1}}\bigg\rvert_{t=0}=2(Q_t,\frac{dQ_t}{dt})\bigg\rvert_{t=0},
\end{align}
where  the inner product $(A,B)$ is defined in Appendix A and we use the fact that $P_{t=0}$ is diagonal in the $|q_k\rangle$ basis. From Eqs. (\ref{7},\ref{B3},\ref{4}) one finds the decoherence time

\begin{equation}
\label{8}
\tau=\frac{1}{2A\sum_{n=0}^{N} (2n+1) \sum_{\{k_l\}} a^2_{k_1,\dots,k_{2n+1}}}.
\end{equation}

Eq. (\ref{8}) can be used to estimate the decoherence time and select the most robust of the fermionic zero modes, listed in Section III, for use in a qbit. Longer lifetimes correspond to zero modes, dominated by contributions of lower orders in $\gamma_i$'s.

\section{The longest decoherence time}

In this section we estimate the longest decoherence time that can be achieved in an interacting system of $2N+1$ Majorana fermions. 
As is illustrated in the case of $N=1$ in Section II, at small $N$ all zero modes can be expected to have comparable decoherence times. Thus, we concentrate on the limit of large $N$.


Based on  Eq. (\ref{8}), one expects  that the local parity fermionic zero mode $\Gamma$ always corresponds to the shortest decoherence time. Thus, any other choice of the zero mode improves the dephasing time. What is the maximal improvement one can achieve? Since the number of the Majorana zero modes is very large at large $N$, one might think that at least one of the very many Majorana modes has the lifetime, much greater than $\tau_\Gamma$. It turns out, however, that such intuition does not work for strongly interacting systems.

We start with an easy case of a noninteracting system, where the qbit Hamiltonian $H$ is quadratic in Majorana operators. Similar to Section II, an orthogonal change of the variables reduces the Hamiltonian  to the canonical form 
$H=i\sum_{n=1}^N a_n\tilde\gamma_{2n-1}\tilde\gamma_{2n}$, where $\tilde\gamma_k$  ($k=0,\dots,2N$) are new Majorana fermions and $\tilde\gamma_0$ does not enter the Hamiltonian. The operator $\tilde\gamma_0$ is a fermionic zero mode. It is a linear combination of the original Majorana fermions $\gamma_k$. 
Consider a qbit, based on the Majorana mode $\tilde\gamma_0$.
Eq. (\ref{8}) shows that the decoherence time of such qbit does not depend on the system size $N$ and equals roughly $1/[2A]$.
For comparison, the decoherence time $\tau_\Gamma$ for the local parity operator $\Gamma=i^N\Pi_{n=0}^{2N}\gamma_n$ scales as $1/N$. Thus, choosing $\tilde\gamma_0$ to build a qbit gives a considerable advantage at large $N$ in agreement 
with Ref. \onlinecite{GS2012}.
Unfortunately, such long lifetime cannot be obtained in a general strongly interacting system.

To see why, we start with a crude simplistic estimate. We will see that it has the same order of magnitude as the rigorous result. The fermionic zero modes, classified in Section III, are linear combinations of products of odd numbers of Majoranas. For our estimate, we will treat the zero modes as random vectors in the linear space of all such products. The total number of linear independent fermionic zero modes is $2^N.$
In a system with $2N+1$ Majorana fermions, one can construct exactly $2^{2N}$ linear independent products $\Pi_k$ of odd numbers ${\rm deg}[\Pi_k]$ of Majorana operators.
 For the sake of a crude estimate, let us make a simplifying assumption that  
up to constant prefactors
a set of $2^N$ linear independent fermionic zero modes  can be chosen from those $2^{2N}$ products,
i.e, monomial functions of Majorana operators. We will denote those linear independent  modes $F_1,\dots,F_{2^N}$.  

Let us estimate the decoherence times for the above $2^N$ modes $F_k$.
Eq. (\ref{8}) shows that this is sufficient to identify the longest possible decoherence time.

A mode $F_k$ is given by a product of ${\rm deg}[F_k]$ Majorana operators. As Eq. (\ref{8}) shows, to find the maximal decoherence time we need to identify the mode $F_{\rm min}$ with the minimum degree ${\rm deg}[F_{\rm min}]=d_{\rm min}$. The dephasing time can then be extracted from the value of $d_{\rm min}$.
Assuming that each $F_n$ is randomly chosen among $\Pi_l$, one expects that $F_{\rm min}$ is typically one of the $K=2^{2N}/2^N=2^N$ operators $\Pi_l$ of the lowest degrees
${\rm deg}[\Pi_l]=d_1\le d_2 \le \dots\le d_K$. 
For a crude estimate we can set $d_{\rm min}\sim d_K$.
 Then Eq. (\ref{8}) yields the dephasing time $\sim 1/[2A d_K]$.
The number $K$ of the operators $\Pi_k$ with ${\rm deg}[\Pi_k]\le d_K$ can be easily found from combinatorics as $K=LC^{d_K}_{2N+1}$, where the binomial coefficient $C_{2N+1}^{d_K}=(2N+1)!/[d_K!(2N+1-d_K)!]$ and $(d_K/2N)^2<L<(d_K+1)/2$. 
Since we are interested in large $N$, the binomial coefficient can be approximated with the Stirling formula and

\begin{equation}
\label{10}
K\approx\frac{L}{\sqrt{2\pi\epsilon N}}\left[\left(\frac{\epsilon}{2}\right)^{\epsilon/2}\left(1-\frac{\epsilon}{2}\right)^{1-\epsilon/2}\right]^{-2N},
\end{equation}
where we define $\epsilon=d_K/N$.
Next, we use the fact that $K=2^N$ to obtain $\epsilon\approx 0.22$. Hence, ${\rm deg}[F_{\rm min}]\sim 0.2N$ and the decoherence time (\ref{8}), $\tau_{\rm max}\sim \frac{5}{2NA}$. This scales as $1/N$ and is only one order of magnitude better than the estimate $\tau_\Gamma\sim 1/[4NA]$ for a qbit, constructed from the local parity operator $\Gamma$ in place of $F_{\rm min}$. 

The above conclusion agrees with the rigorous estimate from Appendix C. The maximal decoherence time depends on the details of the Hamiltonian and can be higher than the result of Appendix C for special Hamiltonians. To formulate our results precisely, we introduce a measure on the ensemble of all possible Hamiltonians $H$ of a system with $2N+1$ Majorana fermions. Specifically, we assume that the measure depends only on the inner product $(H,H)$, defined in Appendix A. The details of the dependence are not important. Then, in the limit of a large $N$, our estimate applies to all Hamiltonians except a set of measure 0. We find that the dephasing time 

\begin{equation}
\label{9}
\tau_{\rm max}<\frac{5}{NA}.
\end{equation}

\section{Interaction with bosons}

So far, our focus has been on a system of fermions. In realistic systems, interaction with Bose degrees of freedom, such as phonons, is possible. The arguments from Sections III and IV easily translate to such situation. We first remove physically unimportant high energy states from the Hilbert space of bosons. Let the dimension of the truncated Hilbert space of bosons be $D_b$. Any operator of the form (\ref{add1}), where the Hamiltonian $H$ includes both Fermi and Bose degrees of freedom, is still a fermionic zero mode; All fermionic zero modes satisfy (\ref{add1}). The number of the linear independent zero modes changes but can be found with essentially the same argument as in Section III. It equals $D_b2^N$. Exactly the same prescription (\ref{interpolation}) as in Section III can be used to construct zero modes, satisfying Eq. (\ref{normalization}). The number of the ways to build a qbit is given by the same expression as before, $2^{N_E-1}$, where the number of the different energy levels expresses now as $N_E=D_b2^N$ for a generic Hamiltonian.

The discussion of dephasing, Section IV, also applies. In particular, only a slight change is necessary in Eq. (\ref{8}). 
The coefficients $a_{\{k_l\}}$ in Eq. (\ref{add2}) should now be understood as Hermitian operators $\hat a_{\{k_l\}}$ in the Hilbert space of bosons.  In order to compute the dephasing time, ${\rm Tr} \hat a^2_{\{k_l\}}/D_b$ should be written instead of $a^2_{\{k_l\}}$ in Eqs. (\ref{add3}) and (\ref{8}) and $D_b2^{N+1}$ instead of $2^{N+1}$ in Eq. (\ref{add3}).

We were unable to extend the rigorous proof of the estimate (\ref{9}) from Appendix C to systems with bosons.  Nevertheless, we expect the limit on the maximal decoherence time (\ref{9}) to hold irrespective of the presence of Bose degrees of freedom. This expectation is supported by the qualitative argument below.

We first note that there are $D_b^2$ linear independent Hermitian operators $\hat A_k$ in the Hilbert space of bosons. They can be selected so that ${\rm Tr}\hat A_i\hat A_j=\delta_{ij}$. Any of the Majorana zero modes,
satisfying Eq. (\ref{normalization}), can be represented in the form 

\begin{equation}
\label{11}
F=\sqrt{D_b}\sum c^{(n)}_{m,\{k_s\}}\hat A_m i^n\Pi_{s=1}^{2n+1}\gamma_{k_s},
\end{equation}
where $k_1<k_2<\dots<k_{2n+1}$, $m=1,\dots,D_b^2$ and $\sum[c^{(n)}_{m,\{k_s\}}]^2=1$. The dephasing time is given by Eq. (\ref{8}) with $c^{(n)}_{m,\{k_s\}}$ in place of $a_{\{k_s\}}$.
For the sake of our qualitative argument we will treat the above  Majorana zero modes (\ref{11}) as random vectors in the space of all possible fermionic operators. The dimension of the latter space is $D_b^2 2^{2N}$.
It will be convenient to set 

\begin{equation}
\label{12}
c^{(n)}_{m,\{k_s\}}=\frac{b^{(n)}_{m,\{k_s\}}}{\sqrt{\sum [b^{(n)}_{m,\{k_s\}}]^2}}.
\end{equation} 
This ansatz takes care about appropriate normalization conditions.
For simplicity we will assume that $b^{(n)}_{m,\{k_s\}}$ are independent Gaussian variables. The results do not depend on their variance and we select the distribution functions of the form 
$P(b^{(n)}_{m,\{k_s\}})\sim \exp[-(b^{(n)}_{m,\{k_s\}})^2D_b^22^{2N}/2]$. This choice implies that $\langle \sum [b^{(n)}_{m,\{k_s\}}]^2\rangle=1$.

Since there are $2^{D_b2^N-1}$ Majorana  modes (\ref{11}), we can neglect the choices of the coefficients $b^{(n)}_{m,\{k_s\}}$ whose joint probability is below $2^{-D_b2^N}$. For example, $\sum [b^{(n)}_{m,\{k_s\}}]^2$, is a random variable with the variance $2/[D_b^2 2^{2N}]$. Thus, we can neglect the probability of the event that $|\sum [b^{(n)}_{m,\{k_s\}}]^2-1|\gg 1/\sqrt{D_b2^N}$. Similar considerations show that we can neglect the probability of such configurations that $\sum_{2n+1<0.1 N} [b^{(n)}_{m,\{k_s\}}]^2\sim 1$. Eq. (\ref{8}) then immediately leads to the estimate (\ref{9}).

In contrast to Appendix C, the above simple argument is not rigorous. In Appendix D, we use a rigorous approach to compare dephasing times for Hamiltonians, quadratic in Majoranas, in the absence and presence of bosonic modes. We find that Bose modes shorten the decoherence time. Thus, we expect that interaction with Bose modes improves the decoherence time in neither weakly nor strongly interacting systems.

\section{Summary}

A topological qbit can be constructed from a fermionic zero mode of a system of $2N+1$ Majorana fermions. We classified such modes $F_k$ for a generic Hamiltonian and found a simple analytical expression for all of them. We also proposed how to design the qbit with the longest decoherence time. One first needs to identify the modes, satisfying the Majorana condition $F_k^2=1$. Second, one should use Eq. (\ref{8}) to estimate the decoherence
time, corresponding to each choice of the zero mode, and select the most robust Majorana operator. In noninteracting systems, the decoherence times vary greatly, depending on the choice of $F_k$. We found that the maximum decoherence time does not depend on the system size
in the absence of interactions. On the other hand,
in strongly interacting systems, the shortest decoherence time $\tau_\Gamma\sim 1/[4NA]$ and the longest decoherence time (\ref{9}) differ by no more than one order of magnitude. The contrast between interacting and noninteracting systems is not surprising. In the absence of interactions, it is possible to find a subsystem that does not feel most excitations in the bath. Such effectively isolated subsystems cannot exist in the presence of strong interactions.

Many-particle systems can exhibit many-body localization \cite{mbl} (MBL) in the presence of weak interactions. Localized states may support topological order \cite{top1,top2} and it was argued that MBL may be used to build robust topological memory \cite{top2} in a large system. This does not conflict with our results in the case of strong interactions, where we find rapid decrease of the decoherence time as a function of $N$. First of all, MBL occurs at weak interactions, where our results do not apply. Even more importantly, a typical Hamiltonian in the space of all Hamiltonians, Appendix C, involves the interaction of all pairs of Majorana modes. This is natural for Majorana fermions, localized in the same vortex, but quite different from what happens in systems with MBL. Such systems have zero measure in the space of the interacting Hamiltonians at $N\rightarrow\infty$.

It was proposed in Ref. \onlinecite{GS2012} that selecting a zero mode with the longest decoherence time is a route to more robust topological quantum computation. Our results show that such strategy is rather limited. The structure of all zero modes, except $\Gamma$, is complicated and sensitive to the system details. It is not obvious how to access them experimentally. Taking into account a relatively small gain in the decoherence time from optimizing $F_k$, we see that the advantages of such optimization are narrow. The only ways to dramatically improve the decay time consist in 1) suppressing the interaction with the bath; 2) reducing the interaction between the fermions that form the qbit or 3) cutting the number of low-energy degrees of freedom by increasing the minigap \cite{sau,tewari}.

\begin{acknowledgments}
We thank Chenjie Wang for useful discussions.
This work was supported by the NSF under Grant No. DMR-1205715.
\end{acknowledgments}

\appendix
\section{Inner product of operators.}

In this appendix we introduce the notion of the inner product of operators, constructed from Majorana fermions. This notion is used in Sections IV and V of the main text and Appendices B and C.

We consider $2N+2$ Majorana fermions $\gamma_0,\dots\gamma_{2N}$ and $f$. They act in the Hilbert space of dimension $D=2^{N+1}$. One can also associate this set of Majoranas with two linear spaces of operators.
The space $L_{\rm even}$ includes all Hermitian operators that express as linear combinations of products of even numbers of the operators $\gamma_i$.  
We will denote such products as $E_{\{k_l\}}=E_{k_1,\dots,k_{2n}}=i^n\Pi_{j=1}^{2n}\gamma_{k_j}$, where $n=0,\dots,N$ and $k_1<k_2<\dots<k_{2n}$. The product of zero $\gamma$'s is defined as $E=1$. Any vector in $L_{\rm even}$ can be represented as $\sum_{n=0}^N\sum_{k_1,\dots,k_{2n}}a_{k_1,\dots,k_{2n}}E_{k_1,\dots,k_{2n}}$.
We prove below that $E_{k_1,\dots,k_{2n}}$ form a basis in $L_{\rm even}$.
Note that the operator $f$ does not enter any of the expressions for  $E_{k_1,\dots,k_{2n}}$. Note also that the Hamiltonian of the qbit is a vector in the space $L_{\rm even}$.

The space of fermionic Hermitian operators $L_{\rm odd}$ is constructed in a similar way. The only difference is that any vector in $L_{\rm odd}$ is a linear combination of the products 
$O_{\{k_l\}}=O_{k_1,\dots,k_{2n+1}}=i^n\gamma_{k_{2n+1}}\Pi_{j=1}^{2n}\gamma_{k_j}$ of odd numbers of the Majorana operators $\gamma_{k_j}$, $k_1<\dots <k_{2n+1}$.

Consider two vectors in the space $L_{\rm even}$: 

\begin{align}
\label{B0}
A=\sum_{n}\sum_{k_1,\dots,k_{2n}}a_{k_1,\dots,k_{2n}}E_{k_1,\dots,k_{2n}}{\rm ~~~and} \nonumber\\
B=\sum_{n}\sum_{k_1,\dots,k_{2n}}b_{k_1,\dots,k_{2n}}E_{k_1,\dots,k_{2n}}. 
\end{align}
We define their inner product as 

\begin{equation}
\label{B1}
( A,B ) ={\rm Tr} AB/2^{N+1}.
\end{equation}
It is easy to see that the product is positive definite and satisfies all other requirements for an inner product in  a Euclidean space. 

Let us check that $(E_{\alpha},E_{\beta})=\delta_{\alpha,\beta}$, where $\delta_{\alpha,\beta}=1$, if the sets of the indices $\alpha$ and $\beta$ are identical, and $\delta_{\alpha,\beta}=0$ otherwise.
This statement is equivalent to the requirement that 

\begin{equation}
\label{B2}
{\rm Tr} E_{k_1,\dots,k_{2n}}=2^{N+1}\delta_{n,0}.
\end{equation}
Proving (\ref{B2}) is easy. It follows immediately after we construct the complex fermions $c_{m}=(\gamma_{k_{2m-1}}+i\gamma_{k_{2m}})/2$ and rewrite $E_{k_1,\dots,k_{2n}}$ as $\Pi_{m=1}^n (2c^\dagger_m c_m-1)$.

We have thus proved that the Hermitian operators $E_{k_1,\dots,k_{2n}}$ form an orthonormal basis in the space $L_{\rm even}$. It then follows that the coefficients $a_{k_1,\dots,k_{2n}}$ and $b_{k_1,\dots,k_{2n}}$ in the expansions (\ref{B0}) of the Hermitian operators $A$ and $B$ must be real. We also get a simple expression for the inner product in terms of the coordinates:
$(A,B)=\sum a_{k_1,\dots,k_{2n}}b_{k_1,\dots,k_{2n}}$.

The definition of the inner product in the space $L_{\rm odd}$ is the same: 
If 

\begin{align}
\label{B4}
A=\sum_{n}\sum_{k_1,\dots,k_{2n+1}}a_{k_1,\dots,k_{2n+1}}O_{k_1,\dots,k_{2n+1}} ~~{\rm and}\nonumber\\
B=\sum_{n}\sum_{k_1,\dots,k_{2n+1}}b_{k_1,\dots,k_{2n+1}}O_{k_1,\dots,k_{2n+1}}
\end{align}
 then

\begin{equation}
\label{B3}
( A,B)={\rm Tr} AB/2^{N+1}=\sum a_{k_1,\dots,k_{2n+1}}b_{k_1,\dots,k_{2n+1}}.
\end{equation}
Just as above, the Hermitian operators $O_{\{k_l\}}$ form an orthonormal basis in the Euclidean space $L_{\rm odd}$ and the coordinates $a_{\{k_l\}}$, $b_{\{k_l\}}$ are always real.

\section{Classification of zero modes.}

We are looking for Hermitian operators $F=F^\dagger$ that commute with the Hamiltonian $H$ and anticommute with the Fermi operator $f$ which creates excitations far away from the Majorana modes $\gamma_i$.

Clearly, 

\begin{equation}
\label{A1}
F=P(H)\Gamma,
\end{equation} 
where $P(H)$ is an arbitrary polynomial of the Hamiltonian and $\Gamma=i^N\Pi_{k=0}^{2N}\gamma_k$ the local parity operator, are integrals of motion that satisfy Fermi statistics. For some Hamiltonians, additional fermionic zero modes are possible. For example, any fermionic operator commutes with the Hamiltonian, if $H=0$. We show below that for a typical Hamiltonian all fermionic zero modes have the form (\ref{A1}) and exactly $2^N$ of such modes are linear independent. Those statements do not hold for a zero-measure set of Hamiltonians only.

It was stated in Ref. \onlinecite{GS2012} that there are exactly $2^N$ fermionic zero modes in the absence of interactions and for infinitesimal interactions. Here we find the number of the zero modes for an arbitrary interaction strength. We also give an explicit formula for all integrals of motion.

We first prove the existence of $2^N$ linear independent integrals of motion of the form (\ref{A1}). This is equivalent to finding $2^N$ linear independent polynomials $P(H)$. We show below that the appropriate polynomials are  $P(H)=H^k$ with $k=0,1,\dots,2^N-1$. 

We start with an example. Let the Hamiltonian be

\begin{equation}
\label{A2}
H=\sum_{n=0}^N\sum_{0<k_1<k_2<\dots<k_n<N+1}u_{k_1,\dots,k_n}\Pi_{j=1}^n\alpha_{k_j} ,
\end{equation}
where $u_{\{k_l\}}$ are constants and $\alpha_k=i\gamma_{2k-1}\gamma_{2k}$. Note that all $\alpha_k$ commute with each other and square to 1. Hence, the multiplicative group $G$ of all possible products of several operators $\alpha_k$ is 
$G=\mathbb
{Z}_2^N$. The identity operator plays the role of the identity element in the group. The Hamiltonian $H$ is an element of the group algebra $R[G]$ of $G$. The same is true for any polynomial $P(H)$. The Abelian group $G$ has ${\rm dim~}G=2^N$ one-dimensional representations $R_k$, $k=1,\dots,2^N$, whose characters are $\pm 1$. Thus, the group can be faithfully represented by $2^N\times 2^N$ diagonal matrices $M(g_s)$, $g_s\in G$, whose nonzero entries $M_{kk}(g_s)$ equal the characters of the elements of $G$ in the representation $R_k$. In such representation, $H$ becomes a diagonal matrix whose diagonal elements $H_{jj}$ are linear combinations $\sum_n\sum_{0<k_1<k_2<\dots<k_n<N+1} \pm u_{k_1,\dots,k_n}$ with a different set of signs in front of the variables $u_{k_1,\dots,k_n}$ for each $j$. For almost any choice of $H$, all $2^N$ entries $H_{jj}$ are different from each other. Consider now an arbitrary linear combination
$\sum_{k=0}^{2^N-1} c_k H^k=Q(H)$, where the degree of the polynomial $Q(x)=\sum c_k x^k$ is $2^N-1$ or lower. $Q(H)$ is a diagonal matrix with the entries $Q(H_{jj})$. They could be simultaneously zero only if all the numbers $H_{jj}$ were the roots of $Q(x)$,  in contradiction with the fundamental theorem of algebra. Hence, $Q(H)\ne 0$. This proves that the operators $H^k$, $k=0,\dots,2^N-1$ are linear independent.

So far the desired property of linear independence was established for a special class of Hamiltonians (\ref{A2}) only. We now demonstrate that the result for this special class  implies the general statement.
Indeed, any Hamiltonian $H$ can be seen as a vector in the space $L_{\rm even}$ (Appendix A) and is defined by its $2^{2N}$ coordinates $a_{\{k_l\}}$ in the basis $E_{\{k_l\}}$ (see Appendix A for details). The components of any power $H^k$ in the same basis are polynomials of $a_{\{k_l\}}$. Let us write the components of the vectors $H^k$, $k=0,\dots,2^N-1$ in the form of a $2^N\times2^{2N}$ matrix $T$. The coordinates of any of the $2^N$ vectors $H^k$ form one row of the matrix. Let us first select an arbitrary Hamiltonian $H_1$ of the form (\ref{A2}). Then the rank of the matrix $T$ is $2^N$ since all of its rows are linear independent. Hence, $T$ has a nonzero minor $M_T$ of size $2^N$.  Let us now consider an arbitrary Hamiltonian $H$. We form the matrix $T$ and compute exactly the same minor as for $H_1$, i.e., select the same columns that form $M_T$ and compute the determinant of the square matrix, formed by those $2^N$ columns. We obtain a polynomial of $a_{\{k_l\}}$. We know that the minor is nonzero for one particular choice of the variables $a_{\{k_l\}}$. Since it is a polynomial, it follows that it is nonzero for almost any other choice. Thus, the rank of the matrix $T$ is $2^N$ almost everywhere and hence the $2^N$ operators $H^k$ are linear independent indeed.

We need to prove the converse statement now: almost all Hamiltonians have no more than $2^N$ fermionic zero modes. We will prove instead that there are no more than $2^N$ bosonic modes. This is enough since multiplying Hermitian bosonic zero modes by $\Gamma$ establishes a one-to-one correspondence between bosonic and fermionic zero modes.

A zero mode $F$ is a Hermitian operator that satisfies the equation $[H,F]=0$. We  consider $F$ as a vector in $L_{\rm even}$ and introduce a linear operator $\tilde H$ from $L_{\rm even}$ to $L_{\rm even}$  such that $\tilde H F=i[H,F]$. Since zero modes form the kernel of $\tilde H$, our goal is to prove that ${\rm dim~ker}\tilde H\le 2^N$ for almost all $H$. 

As above, we first prove that inequality for a particular Hamiltonian
\begin{equation}
\label{A3}
H_0= U \sum_{k=1}^N 2^k \alpha_k,
\end{equation}
where $U\ne 0$.
An arbitrary zero mode is determined by its coefficients in the basis $E_{\{k_l\}}$. It will be convenient to change the notation for the basis vectors. This will help us better exploit the structure of $H_0$. 
We define the following operators: $Z^k_{0,1}=1$, $Z^k_{0,-1}=\alpha_k$, $Z^k_{1,1}=\sqrt{i}\gamma_{2k-1}$, $Z^k_{1,-1}=\sqrt{i}\gamma_{2k}$, $z_0=1$, and $z_1=\sqrt{i}\gamma_0$, where $k=1,\dots, N$.
Any operator $E_{\{k_l\}}$ can be represented as a product $z_{s_0}\Pi_{k=1}^N Z^k_{p_k,s_k}$. We will denote such products as $e_{s_0,p_1,s_1,\dots,p_N,s_N}$.
Let us cut the set of integers $S=\{1,2,\dots,N\}$ into two nonintersecting sets $S_1$ and $S_2$ whose union is $S$. Consider the subspace $L_{S_2}$ of $L_{\rm even}$, spanned by the vectors $e_{s_0,p_1,s_1,\dots,p_N,s_N}$ 
with $p_k=0$ for all $k\in S_1$, $p_k=1$ for all $k\in S_2$, and $s_0=({\rm card~~} S_2) {\rm mod~~} 2$, where ${\rm card}$ means the number of the elements in a set. Any such subspace is an invariant subspace of $\tilde H$.
Thus, in order to find the kernel of $\tilde H$ it is sufficient to find the kernels of its restrictions to the above subspaces.

Let us first consider the subspace $L_{\emptyset}$ that corresponds to $S_2=\emptyset$. This space of dimension $2^N$ is all in the kernel ${\rm ker}\tilde H$. Thus, we need to prove that the restrictions of $\tilde H$ to all other subspaces $L_{S_2}$ have trivial kernels. 

Let us focus on one such subspace $\tilde L=L_{S_2}$, $S_2\ne\emptyset$. It will be convenient to change the notation for the basis vectors in the subspace $\tilde L$ one more time. We define  $e_{s_0,p_1,s_1,\dots,p_N,s_N}=(s_{k_1}, s_{k_2},\dots,s_{k_C})$, where $C$ is the number of the elements of $S_2$ and $k_1>k_2>\dots>k_C$ are the elements of $S_2$. Consider an arbitrary nonzero vector in $\tilde L$, $v=\sum a_{s_{k_1},\dots, s_{k_C}}(s_{k_1},\dots,s_{k_C})$. Let $a_{r_1,\dots,r_C}$ have the maximal absolute value $|m|$ among all the components of $v$. We will now compute the projection of $\tilde H v$ on the vector $(-r_1,r_2,r_3,\dots,r_N)$ (the inner product was defined in Appendix A). We find that the absolute value of the projection is no less than $2|m||U|(2^{k_1}-\sum_{r=2}^C 2^{k_r})>0$. Hence,
$\tilde H v\ne 0$. Since $v$ is an arbitrary vector, we have established that the restriction of $\tilde H$ to $\tilde L$ has a trivial kernel. It follows that ${\rm dim~\rm ker}\tilde H=2^N$ and $H_0$ has exactly $2^N$ bosonic (and hence also fermionic) zero modes.

What about an arbitrary Hamiltonian? The number $K$ of the bosonic (and fermionic) zero modes is determined by the rank of the linear operator $\tilde H$ in the space $L_{\rm even}$: $K=2^{2N}-{\rm rank~}\tilde H$. We want to prove that
${\rm rank~}\tilde H\ge 2^{2N}-2^N$ for almost all Hamiltonians. The latter inequality has been established for $H_0$. Consider the matrix of the operator $\tilde H$, corresponding to $H=H_0$, in the basis $E_{\{k_l\}}$. It has a nonzero minor of the dimension 
$(2^{2N}-2^N)$. The minor is a polynomial of the matrix elements of $\tilde H$. Those matrix elements are, in turn, linear combinations of the coordinates of $H$, if $H$ is interpreted as a vector in $L_{\rm even}$ and expanded as a linear combination of $E_{\{k_l\}}$.
Thus, the minor is a polynomial function of the components of $H$ seen as a vector in $L_{\rm even}$. That polynomial is nonzero at $H=H_0$. It follows that it is nonzero for almost all choices of the components of $H$, i.e., for almost all Hamiltonians. Thus, the rank of $\tilde H$
is almost always at least $2^{2N}-2^N$.
This finishes the proof of the statement of Appendix B.

\section{Structure of zero modes.}

In this Appendix we derive Eq. (\ref{9}). As discussed in Section V, we are only interested in the limit of large $N$.

\subsection{The idea of the argument}

We start with the summary of our argument.

Zero modes are polynomials of $\gamma_k$'s. Long decoherence times correspond to polynomials with large coefficients in front of the terms of low powers in $\gamma_k$ and small coefficients in front of the terms of high powers in $\gamma_k$. We want to prove that for almost all Hamiltonians, all of their fermionic zero modes have decoherence times that scale as $1/N$ and satisfy Eq. (\ref{9}).

In order to make the above statements precise, we introduce a measure on the set of all possible Hamiltonians of a system of $2N+1$ Majorana fermions. This gives a clear meaning to the phrase ``almost all Hamiltonians''. We also need to define what is meant by low and high powers. We introduce a constant $\epsilon<1$. Low powers $p$ satisfy 

\begin{equation}
\label{C1}
p<\epsilon N.
\end{equation}
What is meant by long and short decoherence times also depends on the choice of $\epsilon$: long decoherence times correspond to zero modes whose expressions are dominated by terms of power $<\epsilon N$.
Thus, the choice of $\epsilon$ defines the set of all Hamiltonians $H_{\rm long}$ that possess at least one fermionic zero mode $F_{\rm long}(H_{\rm long})$ with a long decoherence time. We denote the measure of that set of Hamiltonians by $\sigma$ (the measure of the set of all Hamiltonians is normalized to 1). Our goal is to find the smallest $\epsilon$ such that $\sigma$ remains finite in the limit of large $N$. The upper bound  (\ref{9}) on the decoherence time for an arbitrary Hamiltonian outside a set of zero  measure can then be derived from the knowledge of $\epsilon$.

As the first step of the argument, we introduce a family of measure-preserving unitary transformations $U_k$. They transform Hamiltonians into Hamiltonians and zero modes into zero modes. The total number $N_{U}$ of the transformations in the family grows rapidly as a function of $N$.

Next, we consider all Hamiltonians $H_{\rm long}$ that possess at least one fermionic zero mode $F_{\rm long}(H_{\rm long})$ with a long decoherence time. 
We track the fate of the Hamiltonians from that set under the action of each unitary transformation $U_k$. Each pair $(H_{\rm long},F_{\rm long})$ is transformed by $U_k$ into a new Hamiltonian $H_k(H_{\rm long})$ and a fermionic zero mode 
$F_k(F_{\rm long})$ of the new Hamiltonian $H_k(H_{\rm long})$. In general, $F_k(F_{\rm long})$ may have arbitrary coefficients in front of high- and low-power terms. 

At the third step, we sum over $k$ the measures of the sets of the Hamiltonians of the form $H_k(H_{\rm long})$. At large $N$, the sum $\tilde\sigma=\sigma N_U\gg 1$.
This means that those sets intersect and some Hamiltonians $H$ can be represented as $H=H_k(H_{\rm long})$ at multiple choices of $k$ and $H_{\rm long}$. For each of those choices, $F^H_k=F_k(F_{\rm long}(H_{\rm long}))$ is a fermionic zero mode of $H$. Not all of the modes $F_k^H$ are linear independent. We use the structure of the operators $U_k$ to derive the lower bound $\tilde N_F$ on the number 
of the linear independent zero modes $F_k^H$. It assumes the form $\tilde N_F=N_U/r$, where $r$ is a function of $N$ and $\epsilon$, Eq. (\ref{C1}). According to Appendix B, $\tilde N_F\le 2^N$. Hence, $r\ge 2^{-N} N_U$. This yields an inequality for $\epsilon$  whose solution leads to Eq. (\ref{9}).

\subsection{Measure in the space of Hamiltonians}

According to Appendix A, every Hamiltonian $H$ is determined by its coordinates $a_{\{k_l\}}$ in the basis $E_{\{k_l\}}$. We define a volume element in terms of those components and the inner product, introduced in Appendix A:
$dV=f[(H,H)]\Pi_{\{k_l\}}d a_{\{k_l\}}$, where $f(x)$ is an arbitrary function of the inner square of $H$. The only restriction is that the total volume of the space of the Hamiltonians must be finite: $\int dV=1$.

We will not prove that relevant sets are measurable or integrals exist. Such proofs can be deduced from a physically sensible assumption that zero modes depend continuously on the Hamiltonian
for almost all Hamiltonians.

\subsection{Unitary transformations}

Consider the operators

\begin{equation}
\label{C2}
U_{\{k_l\}}=\frac{1+iA_{\{k_l\}}}{\sqrt{2}}=\frac{1}{\sqrt{2}}(1+ i^{M/2+1}\Pi_{l=1}^M \gamma_{k_l}),
\end{equation}
where $M=N$ for even $N$, $M=N+1$ for odd $N$, and $\{k_l\}$ is a shorthand for the set of the indices $k_1<k_2<\dots<k_M$. All such operators are unitary. The total number of different operators $U$ is

\begin{equation}
\label{C3}
N_U=C^M_{2N+1}\approx\frac{2^{2N+1}}{\sqrt{\pi N}}.
\end{equation}

The action of $U_{\{k_l\}}$ on a Hamiltonian $H$ and its fermionic zero mode $\Gamma_\alpha$ is defined by $H\rightarrow H'=UHU^\dagger$, $\Gamma_\alpha\rightarrow \Gamma_\alpha'= U\Gamma_\alpha U^\dagger$.
Clearly, $\Gamma_\alpha'$ is a fermionic zero mode of $H'$. It is also clear that the action of $U_{\{k_l\}}$ preserves the measure, Subsection C.1.


We wish to understand the action of the operators (\ref{C2}) on fermionic zero modes. Each fermionic zero mode can be represented as a linear combination of the vectors $O_{\{k_l\}}$, Eq. (\ref{B4}).
To simplify notations we will use Greek indices with a bar to denote sets of the indices $\{k_l\}$. Thus, we may write $O_{\bar\alpha}$ instead of $O_{k_1,\dots,k_{2n+1}}$ and represent zero modes as 
$\Gamma_{i}=\sum_{\bar\alpha}a_{\bar\alpha}O_{\bar\alpha}$. Let us first consider the action of $U_{\bar\beta}=(1+iA_{\bar\beta})/\sqrt{2}$ on $O_{\bar\alpha}$. There are two possibilities.

1) The index sets $\bar\alpha$ and $\bar\beta$ have an even number of indices in common. In such case, $U_{\bar\beta}$ and $O_{\bar\alpha}$ commute so that $U_{\bar\beta}O_{\bar\alpha}U^\dagger_{\bar\beta}=O_{\bar\alpha}$.

2) The index sets $\bar\alpha$ and $\bar\beta$ have an odd number of indices in common. Then 

\begin{equation}
\label{C3a}
U_{\bar\beta}O_{\bar\alpha}U^\dagger_{\bar\beta}=iA_{\bar\beta}O_{\bar\alpha}=\pm O_{\bar\alpha'}, 
\end{equation}
where $\bar\alpha'=\bar\alpha'(\bar\alpha,\bar\beta)\ne \bar\alpha$.
Note also that $U_{\bar\beta}O_{\bar\alpha'(\bar\alpha,\bar\beta)}U^\dagger_{\bar\beta}=\mp O_{\bar\alpha}$. 

Thus, there are two types of operators $O_{\bar\alpha}$ for each $U_{\bar\beta}$: 1) some operators $O_{\bar\alpha}$ are fixed points of the action of  $U_{\bar\beta}$ and 2) the rest consists of the pairs of operators $O_{\bar\alpha}, O_{\bar\alpha'}$ that  transform into each other by the action of $U_{\bar\beta}$.

Let us fix a number $\epsilon\ll 1$. Consider a fermionic zero mode 

\begin{equation}
\label{C5}
\Gamma_i=\sum_{\bar\alpha}a_{\bar\alpha}O_{\bar\alpha}
\end{equation}
of some Hamiltonian $H$. We introduce the notation $\Gamma_i^{\epsilon}$ for the sum of all monomials of the degrees less than $\epsilon N$
in the above expansion of $\Gamma_i$: 

\begin{equation}
\label{C4}
\Gamma_i^\epsilon=\sum_{\bar\alpha{~\rm contains~ fewer ~than ~}\epsilon N ~{\rm indices}}a_{\bar\alpha}O_{\bar\alpha}.
\end{equation}
The action of $U_{\bar\beta}$ transforms $\Gamma_i^\epsilon$ into the sum $\Gamma_i^{\epsilon,0}+\Gamma_i^{\epsilon,U_{\bar\beta}}$, where $\Gamma_i^{\epsilon,0}$ includes terms of degrees less than $\epsilon N$ in $\gamma_k$'s and
$\Gamma_i^{\epsilon,U_{\bar\beta}}$ combines monomials of degrees between $M-\epsilon N$ and $M+\epsilon N-2$. We define $\Gamma_i^{{\rm rest}, U_{\bar\beta}}$ according to the equation $U_{\bar\beta}\Gamma_i U_{\bar\beta}^\dagger=\Gamma_i^{\epsilon, U_{\bar\beta}}+\Gamma_i^{{\rm rest}, U_{\bar\beta}}$,
and expand $\Gamma_i^{\epsilon, U_{\bar\beta}}=\sum b_{\bar\gamma}O_{\bar\gamma}$ and $\Gamma_i^{{\rm rest}, U_{\bar\beta}}=\sum c_{\bar\gamma}O_{\bar\gamma}$.
It follows from the way how $U_{\bar\beta}$ acts on the operators $O_{\bar\alpha}$, that $b_{\bar\gamma}$ and $c_{\bar\gamma}$ cannot be simultaneously nonzero for any $\bar\gamma$. This statement will be important below. We will refer to it as {\bf Proposition C.3}.

$\Gamma_i^{\epsilon, U_{\bar\beta}}$ depends on $U_{\bar\beta}$ in a complicated way. It will be thus convenient for us to switch from the discussion of the action of an individual unitary operator $U_{\bar\beta}$ on a zero mode $\Gamma_i$ to a discussion of an average action of the whole set of the unitary operators (\ref{C2}) on a given zero mode. Note first that 

\begin{equation}
\label{C6}
(\Gamma_i^{\epsilon,U_{\bar\beta}},\Gamma_i^{\epsilon,U_{\bar\beta}})=\sum_{\substack{\bar\alpha{\rm~has~an~odd~number~of~common~indices~with~}\bar\beta\\\bar\alpha{~\rm contains~ fewer ~than ~}\epsilon N ~{\rm indices}}}a^2_{\bar\alpha},
\end{equation}
where $a_{\bar\alpha}$ are defined in Eq. (\ref{C5}).
Simple combinatorics shows that for each index set $\bar\alpha$ with fewer than $\epsilon N$ indices there are approximately 
$N_U/2$ 
operators $U_{\bar\beta}$  such that the index sets $\bar\alpha$ and $\bar\beta$ have an odd number of indices in common. We use that fact and Eq. (\ref{C6}) to obtain that

\begin{equation}
\label{C7}
\sum_{\bar\beta}(\Gamma_i^{\epsilon,U_{\bar\beta}},\Gamma_i^{\epsilon,U_{\bar\beta}})>cN_U(\Gamma_i^{\epsilon},\Gamma_i^{\epsilon}),
\end{equation}
where the summation runs over all operators $U_{\bar\beta}$ and  the constant $c\approx 1/2$. Upper and lower bounds on $c$ can be easily derived from combinatorics but will not be needed below.

\subsection{Counting zero modes}

The Majorana fermion condition (\ref{normalization}) implies $(O,O)=1$. Thus, we may assume that all zero modes are normalized: $(\Gamma_i,\Gamma_i)=1$.

Let us estimate the decoherence time for a qbit, built using a fermionic zero mode $\Gamma_i$. We will do it in terms of $\Gamma_i^\epsilon$, introduced in the previous subsection. From the normalization condition $(\Gamma_i,\Gamma_i)=1$ one finds:
$(\Gamma_i-\Gamma_i^\epsilon,\Gamma_i-\Gamma_i^\epsilon)=1-(\Gamma^\epsilon_i,\Gamma^\epsilon_i)$. Combining this expression with Eq. (\ref{8}) and the definition of $\Gamma_i^{\epsilon}$, one finds  the decoherence time

\begin{equation}
\label{C-special}
\tau < \frac{1}{2\epsilon N A [1-(\Gamma_i^\epsilon,\Gamma_i^\epsilon)]}.
\end{equation}

 Thus, in the search for a long decoherence time, we need to focus on the zero modes with large $(\Gamma_i^\epsilon,\Gamma_i^\epsilon)$.

At the same time, we are not interested in special cases, corresponding to a zero-measure set of Hamiltonians. Below we fix a positive constant $f<1$ and consider the Hamiltonians $H$ such that each of them has at least one fermionic zero mode $\Gamma_H$, satisfying 

\begin{equation}
\label{C8}
(\Gamma_H^\epsilon,\Gamma_H^\epsilon)>f.
\end{equation}
We assume that $\epsilon$ and $f$ are chosen so that the measure $\sigma(S_f)$ of the set $S_f$ of all such Hamiltonians remains nonzero in the limit of large $N$: $\sigma(S_f)>\mu>0$, where $\mu$ does not depend on $N$.
For each Hamiltonian $H$ in $S_f$ we select exactly one fermionic zero mode $\Gamma_H$, satisfying Eq. (\ref{C8}).

Note that $(\Gamma_H^{\epsilon,U_{\bar\beta}},\Gamma_H^{\epsilon,U_{\bar\beta}})\le (\Gamma_H^\epsilon,\Gamma_H^\epsilon)\le (\Gamma_H,\Gamma_H)=1$. It follows then from Eqs. (\ref{C7},\ref{C8}) that for each $\Gamma_H$, Eq. (\ref{C8}),
there are more than

\begin{equation}
\label{C9}
K_U=fcN_U/2 
\end{equation}
operators $U_{\bar\beta}$ such that 

\begin{equation}
\label{C9a}
(\Gamma_H^{\epsilon,U_{\bar\beta}},\Gamma_H^{\epsilon,U_{\bar\beta}})>cf/2.
\end{equation}
We will denote as $S_H$  the set of all such operators $U_{\bar\beta}$ for a given $\Gamma_H$, $H\in S_f$.
We will also define the sets $S_{U_{\bar\beta}}$ made of all such Hamiltonians $H$ that $U_{\bar\beta}\in S_H$. Each $S_{U_{\bar\beta}}$ is a subset of $S_f$.

Consider now the action of $U_{\bar\beta}$ on the Hamiltonians in the set $S_{U_{\bar\beta}}$. Each Hamiltonian $H\in S_{U_{\bar\beta}}$ is transformed into a new Hamiltonian $H^{U_{\bar\beta}}$. Each $U_{\bar\beta}$ transforms $S_{U_{\bar\beta}}$ into a set $S^{U}_{U_{\bar\beta}}$ of the same measure, 

\begin{equation}
\label{C10}
\sigma(S^U_{U_{\bar\beta}})=\sigma(S_{U_{\bar\beta}}).
\end{equation}

Below we will evaluate the sum of the measures (\ref{C10}) in several ways.
The sum of the measures $\Omega=\sum_{\bar\beta}\sigma(S_{U_{\bar\beta}})$ can be represented as 

\begin{equation}
\label{C11}
\Omega=\sum_{\bar\beta}\sigma(S_{U_{\bar\beta}})=\int dV N(S_H),
\end{equation}
where $\int dV$
means integration over all Hamiltonians with the measure, defined in Subsection C.1, and $N(S_H)$ denotes the number of the elements in the set $S_H$ for $H\in S_f$, $N(S_H)=0$ for $H$ that are not in $S_f$.
The discussion around Eq. (\ref{C9}) shows that all nonzero $N(S_H)$ exceed $K_U$. Hence,

\begin{equation}
\label{C12}
\Omega>K_U\sigma(S_f)>K_U\mu.
\end{equation}
On the other hand, Eq. (\ref{C10}) implies that 

\begin{equation}
\label{C13}
\Omega=\sum_{\bar\beta}\sigma(S^U_{U_{\bar\beta}}).
\end{equation}
Let us introduce a new piece of notations. For each Hamiltonian $\tilde H$, consider all such pairs ($H$,$U_{\bar\beta}$) that $H\in S_f$, $\tilde H=U_{\bar\beta}HU^\dagger_{\bar\beta}$ and $U_{\bar\beta}\in S_H$. 
We will denote the set of such pairs as $\tilde S_{\tilde H}$
and their number as $\tilde N(\tilde H)$.
Note that different pairs ($H$,$U_{\bar\beta}$) in $\tilde S_{\tilde H}$ contain different operators $U_{\bar\beta}$.
 We can now rewrite Eq. (\ref{C13}) as
\begin{equation}
\label{C14}
\Omega=\int dV \tilde N(H).
\end{equation}
Recall that the total measure of the whole space of Hamiltonians is $\int 1 dV=1$.
Hence, a comparison of Eq. (\ref{C14}) with (\ref{C12}) shows that 

\begin{equation}
\label{C15}
\tilde N(H)>K_U\mu
\end{equation}
for the Hamiltonians $H$ from some set of a nonzero measure.

According to Appendix B, a generic Hamiltonian $H_0$ in that set has exactly $2^N$ linear independent fermionic zero modes. On the other hand,  for each pair $(H,U_{\bar\beta})\in \tilde S_{H_0}$ there is a fermionic zero mode $\Gamma_H$ such that: 1) it satisfies Eq. (\ref{C8}) and 2) 

\begin{equation}
\label{C16}
\Gamma_0^{\bar\beta}=U_{\bar\beta}\Gamma_H U_{\bar\beta}^\dagger
\end{equation}
is a zero mode of $H_0$. The number of the zero modes $\Gamma_0^{\bar\beta}$ with different sets of indices $\bar\beta$ is $\tilde N(H_0)$.
At the same time, it is easy to see that $\tilde N(H_0)>K_U\mu>2^N$ at large $N$. Thus, if all $\tilde N(H_0)$ modes $\Gamma_{0}^{\bar\beta}$ were distinct and linear independent we would arrive at a contradiction. In what follows we count the linear independent modes among $\Gamma_{0}^{\bar\beta}$ and use the limit of $2^N$ on their number to estimate $\epsilon$.

Our estimate relies on a geometric lemma and a combinatorial inequality, proven in the next two subsections.

\subsection{Geometric lemma}

Consider $n$ unit vectors $v_1,\dots,v_n$ and another set of $n$ mutually orthogonal unit vectors $e_1,\dots,e_n$ in a $D$-dimensional Euclidean space $E$. Let $(v_k,e_k)^2>g$ for each $k$. Then the dimension $d$ of the linear space $V$, spanned by the $n$ vectors $v_k$, is greater than $gn$. 

{\bf Proof.} The projection $P_V(e_k)$ of the vector $e_k$ onto the space $V$ cannot be shorter than the absolute value of the inner product of $e_k$ with an arbitrary unit vector in $V$. Hence, $P^2_V(e_k)\ge (v_k,e_k)^2=g$. Since the unit vectors $e_k$, $k=1,\dots,n$ are mutually orthogonal, we can consider them as a part of the orthonormal basis $e_1,\dots,e_n,e_{n+1},\dots,e_D$ in the space $E$. Clearly, $S=\sum_{i=1}^D P_V^2(e_i)\ge \sum_{i=1}^n P^2_V(e_i)>gn$.
Let $w_1,\dots,w_d$ be an orthonormal basis in $V$. Then $S=\sum_{i,j}(e_i,w_j)^2=\sum_{j=1}^d(w_j,w_j)=d$. It follows that $d>gn$.

\subsection{Combinatorial inequality} 

Consider two sets $\bar\alpha$ and $\bar\beta$ of $M$ indices $k^\alpha_1<k^\alpha_2<\dots<k^\alpha_M$ and $M$ indices $k^\beta_1<k^\beta_2<\dots<k^\beta_M$, assuming values between 0 to $2N$. The constant $M$ was defined in the beginning of Subsection C.3. Let us define the {\it overlap} $o(\bar\alpha,\bar\beta)$ of $\bar\alpha$ and $\bar\beta$ as the number of the common indices in the sets $\{k^\alpha_1,k^\alpha_2,\dots,k^\alpha_M\}$ and $\{k^\beta_1,k^\beta_2,\dots,k^\beta_M\}$.
For a given $\bar\alpha$, we wish to estimate the number $R$ of the sets of indices $\bar\beta$ whose overlap with $\bar\alpha$ exceeds $(M-\epsilon N)$.

Consider an arbitrary subset $[\bar\alpha]_{\rm short}$ of $(M-\epsilon N)$ indices in $\bar\alpha$. Let us count all sets $\bar\gamma$ which contain the same subset $[\bar\alpha]_{\rm short}$. Next, let us add the resulting numbers for every choice of $[\bar\alpha]_{\rm short}$. This way we will count every $\bar\gamma$ with $o(\bar\alpha,\bar\gamma)>(M-\epsilon N)$ at least once. Thus, we will get an upper estimate for $R$. One can choose $[\bar\alpha]_{\rm short}$ in $C_M^{M-\epsilon N}$ ways.  There are $C_{2N+1-M+\epsilon N}^{\epsilon N}$ ways to complement $[\bar\alpha]_{\rm short}$ by $\epsilon N$ additional indices to make a set of $M$
different indices.  Thus, we obtain the inequality $R<C_M^{M-\epsilon N}C_{2N+1-M+\epsilon N}^{\epsilon N}$. 
Using the inequality
$$
\frac{(2N+1-M+\epsilon N)!}{(M-\epsilon N)!}<(2N+1-M+\epsilon N)^{2N+1-2M+2\epsilon N}
$$
and the Stirling formula, we estimate

\begin{equation}
\label{C17}
R<\frac{C}{\epsilon N}\left(\frac{[1+\epsilon]e}{\epsilon}\right)^{2\epsilon N},
\end{equation}
where the constant $C$ does not depend on $N$ and $\epsilon$.

\subsection{Linear independent zero modes}

We now come back to the end of Subsection C.4 and determine the number of linear independent zero modes among the modes $\Gamma_0^{\bar\beta}$ of the Hamiltonian $H_0$. Recall that Eq. (\ref{C16}) establishes a correspondence between the modes
$\Gamma_0^{\bar\beta}$ and the zero modes $\Gamma_H$ of the Hamiltonians $H$ such that $H_0=U_{\bar\beta} H U^\dagger_{\bar\beta}$, $(H,U_{\bar\beta})\in \tilde S_{H_0}$. 


We now observe that the operators $\Gamma_H^{\epsilon,U_{\bar\beta}}$ satisfy Eq. (\ref{C9a}).
Let us introduce the set $S_e$ of the unit vectors $e_{\bar\beta}=\Gamma_H^{\epsilon,U_{\bar\beta}}/\sqrt{(\Gamma_H^{\epsilon,U_{\bar\beta}},\Gamma_H^{\epsilon,U_{\bar\beta}})}$ in the space $L_{\rm odd}$.
The number $N(S_e)$ of the vectors in  $S_e$ satisfies Eq. (\ref{C15}), i.e.,

\begin{equation}
\label{C20}
N(S_e)>K_U\mu=fc\mu N_U/2.
\end{equation}
Note that $N(S_e)$ counts different index sets $\bar\beta$ and it may happen that some $e_{\bar\beta}=e_{\bar\gamma}$ at $\bar\beta\ne\bar\gamma$.
  Eq. (\ref{C9a}) can be rewritten as

\begin{equation}
\label{C18}
(e_{\bar\beta},\Gamma_H^{\epsilon, U_{\bar\beta}})^2>cf/2.
\end{equation}
Proposition C.3 implies now that

\begin{equation}
\label{C19}
(e_{\bar\beta},\Gamma_0^{\bar\beta})^2>cf/2.
\end{equation}

At this point our tactics becomes obvious: In order to find the number of linear independent modes $\Gamma_0^{\bar\beta}$, we need to use geometric lemma C.5 with $\Gamma_0^{\bar\beta}$
in place of $v_k$ and
 $e_{\bar\beta}$ in place of $e_k$. However, a difficulty emerges: In general, the unit vectors $e_{\bar\beta}$ are not mutually orthogonal. That is where the combinatorial inequality (\ref{C17}) is going to help. 

Any vector $e_{\bar\beta}$ is a linear combination of some operators of the form $O^{\epsilon,\bar\beta}_{t}= U_{\bar\beta}O^{\epsilon}_{t}U^\dagger_{\bar\beta}$, where the operators $O^{\epsilon}_{t}$
satisfy two conditions: 

1) $O^{\epsilon,\bar\beta}_{t}\ne O^{\epsilon}_{t}$;

2) $O^{\epsilon}_{t}$ are the vectors $O_{\{ k_l\}}$, Appendix A, with  fewer than $\epsilon N$ indices.\\

We can be sure that $(e_{\bar\beta},e_{\bar\gamma})=0$, if there is no overlap between the sets of all operators $O^{\epsilon,\bar\beta}_{t}$ and all operators $O^{\epsilon,\bar\gamma}_{t}$. Using Eq. (\ref{C3a}) one can show that this is guaranteed to occur,
if $o(\bar\beta,\bar\gamma)\le M-\epsilon N$, where the overlap function $o$ is defined in Subsection C.6. We now select a  vector $e_{\bar\beta_1}$. There are no more than $R$, Eq. (\ref{C17}), operators $e_{\bar\gamma}$ in $S_e$ such that $o(\bar\beta_1,\bar\gamma)> M-\epsilon N$, $\bar\gamma\ne\bar\beta_1$.
Remove all such vectors $e_{\bar\gamma}$ from $S_e$. All remaining vectors $e_{\bar\alpha}\in S_e$, $\bar\alpha\ne \bar\beta_1$ are orthogonal to $e_{\bar\beta_1}$. We next select an arbitrary 
vector $e_{\bar\beta_2}\ne e_{\bar\beta_1}$. By removing no more than $R$ additional vectors from $S_e$ we guarantee that all remaining vectors are orthogonal to $e_{\bar\beta_2}$. We then select an arbitrary remaining vector $e_{\bar\beta_3}\ne e_{\bar\beta_{1,2}}$ and continue in the same spirit until only selected vectors $e_{\bar\beta_k}$ remain in $S_e$. Clearly, $e_{\bar\beta_k}$ form an orthonormal basis. Since we started with at least $K_U\mu$ elements in $S_e$, we end with at least $K_U\mu/R$ orthogonal vectors $e_{\bar\beta_k}$.

Finally, we use geometric lemma C.5 and Eq. (\ref{C19}) to estimate the number $\tilde N_F$ of linear independent zero modes among the operators $\Gamma_0^{\bar\beta}$. 
$\Gamma_0^{\bar\beta_k}$ play the role of the vectors $v_k$ in the lemma and $e_{\bar\beta_k}$ play the role of the vectors $e_k$. There may be additional linear independent modes $\Gamma_0^{\bar\beta}$ with $\bar\beta\ne\bar\beta_1,\dots,\bar\beta_k$. $H_0$ may also have zero modes that do not assume the form $\Gamma_0^{\bar\beta}$. Thus, the total number of linear independent modes $N_F\ge\tilde N_F$.
We find

\begin{equation}
\label{C21}
N_F>\frac{cfK_U\mu}{2R}=\frac{(cf)^2\mu}{4}\frac{N_U}{R},
\end{equation}
where $4R/[\mu(cf)^2]$ plays the role of $r$ from Section C.1.

\subsection{The lowest decoherence rate}

Only one step is left: we will use Eq. (\ref{C21}) to estimate $\epsilon$. Combining Eqs. (\ref{C3},\ref{C17},\ref{C21}) and using the fact that $N_F=2^N$, one finds

\begin{equation}
\label{C22}
2^N>2^{2N }\frac{(cf)^2\mu\epsilon\sqrt{N}}{2C\sqrt{\pi}}\left(\frac{\epsilon}{[1+\epsilon]e}\right)^{2\epsilon N}.
\end{equation}
In the limit of large $N$, Eq. (\ref{C22}) reduces to $\ln 2<2\epsilon [1+\ln\frac{1+\epsilon}{\epsilon}]$. The solution is $\epsilon>0.103$.
In other words, for any $\epsilon<0.103$ and for almost all Hamiltonians $H$, all fermionic zero modes $\Gamma_H$ have vanishing $(\Gamma^\epsilon_H,\Gamma^\epsilon_H)$. We now substitute $\epsilon=0.1$ into Eq. (\ref{C-special}) and obtain Eq. (\ref{9}). Finally, we must mention that some of the modes $\Gamma_H$ do not satisfy the Majorana fermion condition $\Gamma_H^2=1$, Section III, and hence Eq. (\ref{9}) is meaningless for such modes.

\section{Bosonic degrees of freedom in contact with weakly interacting fermions}

In this Appendix we investigate the effect of bosonic degrees of freedom on the dephasing time (\ref{8}). 
We address only the case of weakly interacting fermions below, i.e., we assume that the Hamiltonian is quadratic in Majorana operators.
We also assume that the number $2N+1$ of the fermions is large. No assumptions are made about the dimension $D_b$ of the Hilbert space of bosons (certainly, $D_b>1$).

In the absence of Bose degrees of freedom, any quadratic Hamiltonian has an integral of motion, linear in Majorana operators. We show below that this is no longer the case in the presence of bosons. According to Eq. (\ref{8}) this means that the interaction with Bose degrees of freedom shortens the dephasing time.

We demonstrate this by working with a particular Hamiltonian:

\begin{equation}
\label{D1}
H=i\hat A\sum_{n=0}^{N-1}\gamma_{2n}\gamma_{2n+1}+i\hat B\sum_{n=1}^{N}\gamma_{2n-1}\gamma_{2n},
\end{equation}
where $\hat A$ and $\hat B$ are Hermitian $D_b\times D_b$ matrices acting in the Hilbert space of the bosons. We select the basis in which $\hat B$ is diagonal and assume that its eigenvalues $b_k$, $k=1,\dots,D_b$ are nondegenerate.
We also assume that all matrix elements $A_{ij}$ of the operator $\hat A$ are nonzero in that basis. 

Does a linear zero mode $\Gamma_L=\sum \hat C_k\gamma_k$ exist? Here $\hat C_k$ are operators in the Hilbert space of bosons. In order to answer the question we compute the commutator $[H,\Gamma_L]$. It must be zero for 
$\Gamma_L$ to be an integral of motion. The commutator contains one- and three-fermion contributions. In particular, for each $k>3$, the commutator contains contributions $X_1=i\gamma_1\gamma_2\gamma_k[\hat B,\hat C_k]$ and
$X_2=i\gamma_0\gamma_1\gamma_k[\hat A,\hat C_k]$. For each $k<2N-3$, there are contributions $X_3=i\gamma_k\gamma_{2N-1}\gamma_{2N}[\hat B,\hat C_k]$ and $X_4=i\gamma_k\gamma_{2N-2}\gamma_{2N-1}[\hat A,\hat C_k]$. Each of those contributions $X_i=0$. Hence, $[\hat B,C_k]=[\hat A,C_k]=0$. This is only possible, if each $\hat C_k$ reduces to a $c$
-number: $\hat C_k=c_k\delta_{ij}$. Thus, $\Gamma_L=\sum c_k\gamma_k$ and

\begin{align}
\label{D2}
[H,\Gamma_L]=2i\hat A\sum_{n=0}^{N-1} (c_{2n+1}\gamma_{2n}-c_{2n}\gamma_{2n+1})\nonumber\\
+2i\hat B\sum_{n=0}^{N-1}(c_{2n+2}\gamma_{2n+1}-2c_{2n+1}\gamma_{2n+2})=0.
\end{align}
This means that $\hat Ac_{2n+1}=\hat B c_{2n-1}$ and $\hat A c_{2n}=\hat B c_{2n+2}$ and hence all $c_k=0$. 

We find that no zero mode, linear in Majoranas, exists for the Hamiltonian (\ref{D1}). It may exist at other choices of the Hamiltonian. Still, a general conclusion about the relaxation time being $\tau=(2N+1)\tau_\Gamma$ no longer holds in the presence of bosons.


\begin{thebibliography}{99}
\bibitem{footnote1} A crude estimate of the interaction between fermions, localized in a vortex core, can be obtained from the following model. We describe the core as a disk of normal metal, separated by an infinite barrier from the rest of the system. The radius of the disk $r\sim \hbar v_F/\Delta$, where $v_F$ is the Fermi velocity and $\Delta$ the bulk gap. Bogoliubov quasiparticles do not participate in the long-range Coulomb interaction. Their short-range screened Coulomb interaction has the radius of $l\sim 1/k_F$, where $k_F$ is the Fermi momentum. Its strength $U\sim\hbar v_F k_F$. Hence, the interaction energy of two quasiparticles can be estimated as $U (l/r)^2$ and is comparable with the minigap. Additional contributions to the interaction come from the BCS attraction between electrons.
\end{thebibliography}

\end{document}